\documentclass{emulateapj}
\usepackage{color, hyperref, epsfig}
\usepackage{apjfonts, natbib}

\newcommand{\oiii}{[O\,{\footnotesize III}]}
\newcommand{\loiii}{$L_{\rm \oiii}$}

\newcommand{\lum}{erg~s\ensuremath{^{-1}}}
\newcommand{\lbol}{\ensuremath{L\mathrm{_{bol}}}}
\newcommand{\msun}{\ensuremath{M_{\odot}}}

\newcommand{\wise}{\emph{WISE}}
\newcommand{\neowise}{\emph{NEOWISE}}
\newcommand{\tsub}{\ensuremath{T_\mathrm{sub}}}
\newcommand{\tdust}{\ensuremath{T_\mathrm{dust}}}
\newcommand{\tdustt}{\ensuremath{T_\mathrm{dust,2}}}
\newcommand{\rsub}{\ensuremath{R_\mathrm{sub}}}
\begin{document}

\title{
The \wise\ Detection of an Infrared Echo in Tidal Disruption Event ASASSN-14li
}
\author{{Ning~Jiang\altaffilmark{1}},  Liming~Dou\altaffilmark{1}, 
Tinggui~Wang\altaffilmark{1}, Chenwei~Yang\altaffilmark{1}, Jianwei~Lyu\altaffilmark{2},
Hongyan Zhou\altaffilmark{1,3}}
\altaffiltext{1}{Key laboratory for Research in Galaxies and Cosmology,
Department of Astronomy, The University of Science and Technology of China,
Chinese Academy of Sciences, Hefei, Anhui 230026, China; ~jnac@ustc.edu.cn}
\altaffiltext{2}{Steward Observatory, University of Arizona, 933 North Cherry Avenue, Tucson, AZ 85721, USA}
\altaffiltext{3}{Polar Research Institute of China,451 Jinqiao Road, Pudong,
                 Shanghai 200136, China}
%\altaffiltext{2}{Department of Physics, Anhui Normal University, Wuhu, Anhui, 241000, China}
%%%%%%%%%%%%%%%%%%%%%%%%%%%%%%%%%%%%%%%%%%%%%%%%%%%%%%%%%%%%%%%%%%%%%%%%%%%%%%%%

\begin{abstract}

We report the detection of a significant infrared variability of the nearest 
tidal disruption event (TDE) ASASSN-14li using $\emph{Wide-field Infrared Survey Explorer}$ 
and newly released $\emph{Near-Earth Object WISE Reactivation}$ data. 
In comparison with the quiescent state, the infrared flux is 
brightened by 0.12 and 0.16 magnitude in the W1 ($3.4\mu$m) and W2 ($4.6\mu$m) 
bands at 36 days after 
the optical discovery (or $\sim110$ days after the peak disruption date).
The flux excess is still detectable $\sim170$ more days later. 
Assuming that the flare-like infrared emission is from the dust around the black hole, 
its blackbody temperature is estimated to be $\sim2.1\times10^3$~K, 
slightly higher than the dust sublimation temperature,
indicating that the dust is likely located close to the dust sublimation radius.
The equilibrium between the heating and radiation of the dust claims a 
bolometric luminosity of $\sim10^{43}-10^{45}$~\lum, comparable with the 
observed peak luminosity. This result has for the first time confirmed 
the detection of infrared emission from the dust echoes of TDEs.

%In comparison with the quiescent state,  
%best explained by echo

\end{abstract}


\section{Introduction}

Evidence has mounted that supermassive black holes (SMBHs) with masses of
$10^{6-10}$~\msun\ are present in most (possibly all) galaxies with massive bulges,
and the BH mass correlates tightly with various bulge properties
(see Kormendy \& Ho 2013 for a review). 
When a star orbits too close to the black hole, the tidal force exceeds the star's 
self-gravity and hence the star gets tidally disrupted. 
In these so-called tidal disruption events (TDEs), roughly half of the mass of 
the star may be ejected while the rest of the stellar material is accreted on to 
the black hole, producing a luminous flare of electromagnetic radiation (Rees 1988). 
Their emission peaks in the UV or soft X-rays, with a characteristic $t^{-5/3}$
decline on the timescale of months to years (Evans \& Kochanek 1989; Phinney 1989).
A number of TDE candidates were discovered by flares in the X-ray, UV, optical
and by peculiar superstrong coronal lines in the optical spectra
(see the review by Komossa 2015).
Theoretically, the event rate is estimated to be $10^{-4}$ to $10^{-5}$ 
$\rm galaxy^{-1} yr^{-1}$ (Rees 1988; Magorrian \& Tremaine 1999),
and might be highest in nucleated dwarf galaxies (Wang \& Merritt 2004, Stone \& Metzger 2016).

One of the most promising scopes of TDEs is to probe the circum-nuclear 
interstellar medium (ISM) of inactive galaxies, i.e., by detecting the echoes of the 
gas photoionized by the flare (e.g., Komossa et al. 2008; Wang et al. 2011, 2012;
Yang et al. 2013; Arcavi et al. 2014). 
If the medium is dusty, most of the UV-optical photons will 
be absorbed by dust grains in the vicinity of the BH and re-radiated in the infrared (IR). 
Lu et al. (2016) show that the dust emission peaks at 3-10~$\mu$m and has 
typical luminosities between $10^{42}$ and $10^{43}$~\lum,
which is detectable by current-generation telescopes.
Nevertheless, none of such detections have been reported as of today.
Taking advantage of data from the $\emph{Wide-field Infrared Survey Explorer}$
(\wise; Wright et al. 2010) and newly released data from the 
Near-Earth Object \wise\ Reactivation mission (\neowise-R; Mainzer et al. 2014), 
we have studied the IR light curve of the nearby TDE ASASSN-14li and detected 
significant variability, that is probably the first confirmation of dust echoes from TDEs.

ASASSN-14li was originally discovered on 2014 November 22 (modified Julian day MJD 56,983) 
by the All-Sky Automated Survey for Supernovae (ASASSN, Jose et al. 2014). 
The discovery was triggered by an optical flare coincident with the center of
the galaxy PGC043234 within 0.04\arcsec, prompting it to be 
a potential TDE flare. Later, the $Swift$ X-ray observations established 
a new source at the same location.
%A fit to the UV light curve ($L\propto t^{-5/3}$) gives a disruption MJD of 56,948, 
%that's 35 days before the discovery (Miller et al. 2015).
This galaxy is located at $z=0.0206$, making ASASSN-14li the closest TDE discovered in 
over ten years. The fitting of the light curves using a Monte Carlo approach found that
the accretion rate first exceeded the Eddington ratio in 2014 July,
and the peak luminosity occurred around September 6, which is
two months and a half before the discovery (Alexander et al. 2016). 
Since its discovery, ASASSN-14li has attracted much attentions and 
multiwavelength follow-up monitorings were carried out immediately. 
Light curves and spectra in the optical and near UV (Holoien et al. 2016; Cenko et al. 2016), 
soft X-rays (Miller et al. 2015), and radio (Alexander et al. 2016; van Velzen et al. 2016a) 
have all been obtained.

\begin{figure*}[t]
\centering
\includegraphics[width=16cm]{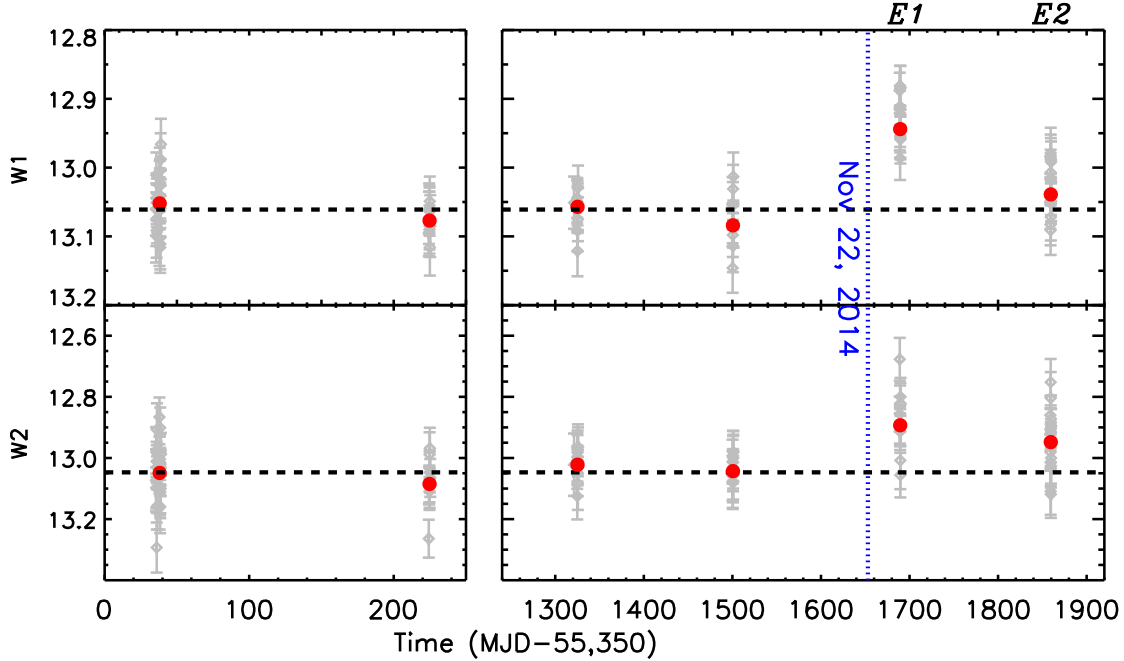}
\caption{The \wise\ infrared light curves of ASASSN-14li in the W1 and W2 band.
Left panel: ALLWISE data; right: \neowise-R data.
The red solid circles denote the median value in each epoch while the single
exposures are in gray.
The uncertainties of these median magnitudes in each epoch are $\lesssim0.01$
in W1 and $\lesssim0.03$ in W2.
The black dashed line is the median magnitude of the first four epochs,
representing the quiescent state.
The blue dotted line denotes the date when ASASSN-14li was discovered
in ASASSN survey.
}
\label{lc}
\end{figure*}

\begin{figure*}
\centering
\includegraphics[width=16cm]{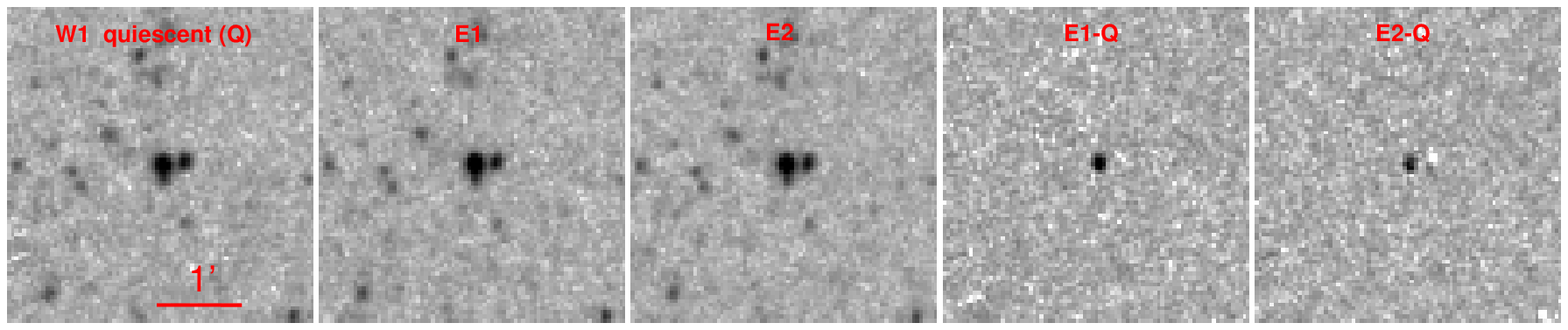}
\hfil
\includegraphics[width=16cm]{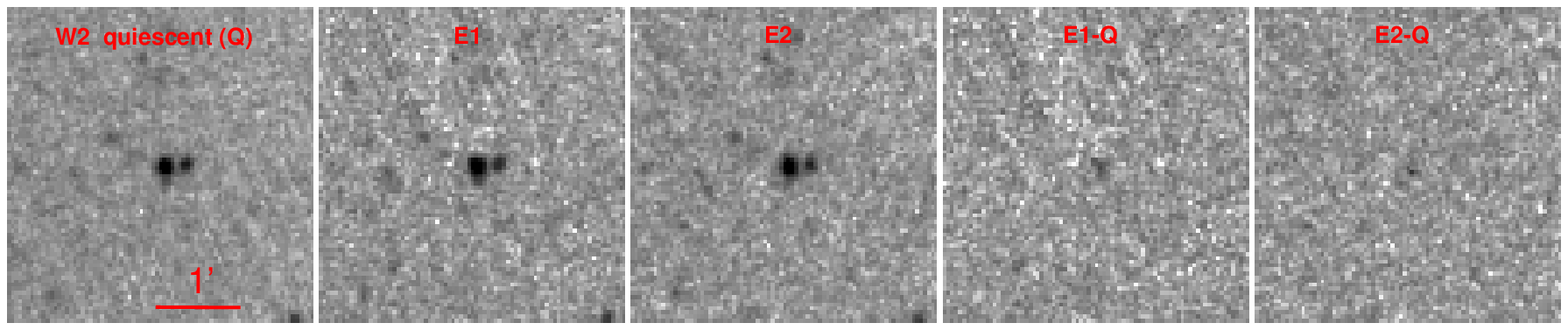}
\caption{
The W1 and W2 images in different epochs.
From left to right: the coadded image of the quiescent state (Q), E1, E2,
the residual image of E1-Q; the residual image of E2-Q.
}
\label{img}
\end{figure*}

However, we still lack the knowledge of the infrared properties of ASASSN-14li,
which is the reprocessed emission from dust located around the BH.
Fortunately, the \neowise-R data that was released on 2016 Mar 23
\footnote{http://wise2.ipac.caltech.edu/docs/release/neowise/},
offers us a good opportunity to probe the IR emission associated with ASASSN-14li.
The \wise\ has performed a full-sky imaging survey in four broad mid-infrared 
bandpasses centered at 3.4, 4.6, 12 and 22~$\mu$m (labeled W1-W4)
from 2010 February to August. 
The solid hydrogen cryogen used to cool the W3 and W4 instrumentation 
was depleted later and it was placed in hibernation in 2011 February. 
\wise\ was reactivated and renamed \neowise-R since 2013 October, using only W1 and W2, 
to hunt for asteroids that could pose as impact hazard to the Earth. 

We assume a cosmology with $H_{0} =70$ km~s$^{-1}$~Mpc$^{-1}$, $\Omega_{m} = 
0.3$, and $\Omega_{\Lambda} = 0.7$.  At a redshift of $z=0.0206$, ASASSN-14li
has a luminosity distance of 89.5~Mpc.

%\section{Data Analysis and Results}

\section{\wise\ Light Curve of ASASSN-14li}

ASASSN-14li is bright enough to have been detected in single exposures throughout
the \wise\ mission. We have collected both cryogenic and post-cryogenic multiepoch
photometry from the AllWISE~\footnote{http://wise2.ipac.caltech.edu/docs/release/allwise/}
and the most recent \neowise-R data releases.
The light curves of ASASSN-14li in W1 and W2 bands have a baseline of 5 years
(MJD 55,365-57,190) and are presented in Figure~\ref{lc}.
The data points are mainly distributed over six epochs at intervals of
about six months with the latest data taken in 2015 June.
There is a large gap ($\sim10^3$ days) between epoch 2 and epoch 3
due to the hibernation of \wise\ between 2011 February and 2013 October.
Each epoch has 10-24 single-exposure observations,
provided by the unique-designed observational
cadence of \wise, which is well suited for studying rapid variability
(e.g. Jiang et al. 2012).
We do not include W3 and W4 data that are restricted to the
cryogenic period of the \wise\ mission.

\begin{deluxetable*}{cccccccccccc}
\tabletypesize{\small}
\setlength{\tabcolsep}{0.04in}
\tablecaption{{\it {\it WISE}}\ Data
\label{tbl:data}}
\tablewidth{0pt}
\tablehead{
\colhead{Epoch} & \colhead{MJD} & \colhead{W1} & \colhead{W2} & \colhead{F1} & \colhead{F2} &
\colhead{\tdust} & \colhead{$L_{\rm IR}$} & \colhead{$L_{\rm bol}$} &
\colhead{\tdustt} & \colhead{$L_{\rm IR,2}$} & \colhead{$L_{\rm bol,2}$}  \\
\colhead{(1)} & \colhead{(2)}  & \colhead{(3)}  & \colhead{(4)} &
\colhead{(5)} & \colhead{(6)}  & \colhead{(7)}  & \colhead{(8)} &
\colhead{(9)} & \colhead{(10)} & \colhead{(11)} & \colhead{(12)} }
\startdata
Q  & 55365.546-56831.170 &  13.06 & 13.05 &  ...   & ...  & ...   &  ...  & ...  & ... & ...  & ...         \\
E1 & 57018.881-57019.867 &  12.94 & 12.89 &  0.21  & 0.16 & 2091  &  41.6 & 45.6 & 904 & 41.3 & 42.9(44.0)  \\
E2 & 57188.984-57189.968 &  13.04 & 12.95 &  0.04  & 0.10 & 588   &  41.0 & 45.0 & 444 & 40.8 & 41.9(43.0)
\enddata
\tablecomments{
Column~(1): observational epochs.
Column~(2): modified Julian Date.
Column~(3): median W1 magnitude.
Column~(4): median W2 magnitude.
Column~(5): enhanced flux in W1, in unit of mJy.
Column~(6): enhanced flux in W2, in unit of mJy.
Column~(7): dust temperature assuming blackbody radiation.
Column~(8): the logarithmic integrated dust IR luminosity.
Column~(9): bolometric luminosity required to heat the dust.
Column~(10): dust temperature considering the dust absorption efficiency ($Q_{\rm abs}\propto \nu^2$).
Column~(11)-(12): same as Column~(8)-(9) while with the case as described in Column~(10).
}
\label{wise}
\end{deluxetable*}

First, no reliable variability is detected within each epoch. Second, we have not
found any obvious flux variations between the first four epochs (MJD-55350$<$1600).
However, we can see a significant flux increase during the fifth epoch (MJD$\approx$57,019),
that is 36 days after the TDE discovery time (Jose et al. 2014).
We take the former four epochs as the quiescent state of the IR emission from the host
in light of the fact that the magnitudes remain almost constant.
For convenience, we denote the quiescent state as "Q" and the two epochs after
the discovery of the TDE as "E1" and "E2" (see Figure~\ref{lc}).
The median magnitudes in the W1 and the W2 bands at E1 are increased
by 0.12 and 0.16 mag, respectively.
The sudden flare-like characteristic, coordinated with the optical/UV/X-ray
flare, makes it probably the first detection of TDE in the infrared.
The flux excess at E2 becomes weaker, yet still detectable
in comparison with the quiescent state (see Table~\ref{wise}).

To exclude any possible systematic biases that make a spurious variability signal,
we have examined the photometric zeropoint offsets between the different frames
by employing nearby stars as "standards".
The zeropoint errors are found to be less than 0.02 mag, which is much lower than
the variation between E1 and Q (see also Jiang et al. 2012).
We have also constructed coadded images in the three states (Q, E1, E2), respectively.
After subtracting the quiescent state, clear flux excesses are shown in the W1 images,
at both E1 and E2; the excesses in the W2 bands are also visible,
albeit  weaker due to the noisy background (see Figure~\ref{img}).

%\section{Derived Physical Parameters of the Dust Emission}

\section{Discussions and Conclusions}

The increased IR emission associated with ASASSN-14li can be naturally attributed to 
the dust-reprocessed light of the optical-UV photons produced by the TDE.
Because the dust is located at some distance from the BH, the IR variability
is delayed with respect to the UV-optical emission
by the light-travel time from the BH to the dust.
As a rough estimation, the time delay after the maximum luminosity 
($\Delta t\sim110$~days) leads to a distance of $\sim0.09$pc, 
or $\sim3\times10^5$ Schwarzschild radius ($r_g$)
given a black hole mass of $10^{6.5}$~\msun.

\begin{figure*}
\centering
\includegraphics[width=16cm]{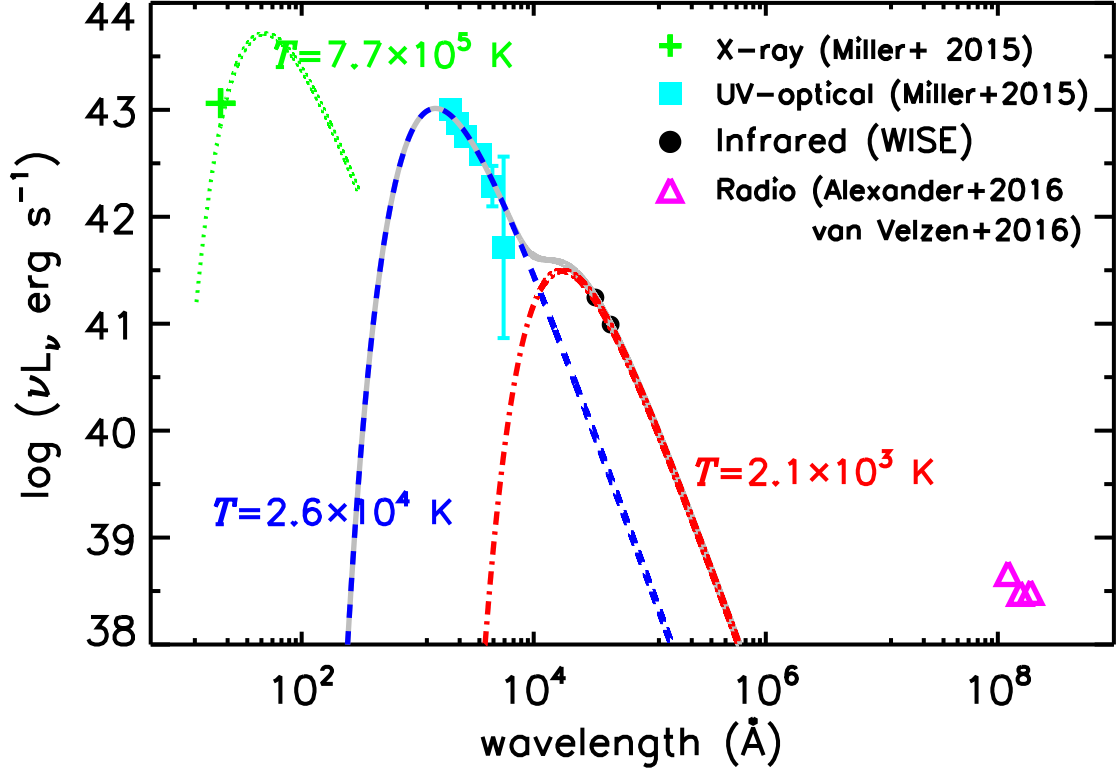}
\caption{
Broadband SED of ASASSN-14li at E1 (MJD$\approx$57019, see Figure~\ref{lc}).
The X-ray, UV and optical data are adapted from Fig. 1 of Miller et al. (2015),
the radio data are drawn from Table S1 of van Velzen et al. (2016a) and
Table 1 of Alexander et al. (2016).
The infrared data are coverted from our measurement.
The blue dashed line and red dotted-dashed line are the two-blackbody fitting to
the UV-optical-infrared data, which have been already subtracted from
the host emission before the TDE.
}
\label{sed}
\end{figure*}

Having been monitored in multiple wavelengths (soft X-rays, UV, optical, and radio) 
after its discovery, ASASSN-14li possesses an exceptionally rich data set up to now.
We have collected all available data reported in the literatures,
which are quasi-simultaneous with the \wise\ observations. 
With the newly added \wise\ IR photometry, we have constructed a
host-subtracted spectral energy distribution (SED) with the most complete 
wavelength coverage among all known TDEs. 
Figure~\ref{sed} shows the broadband SED of ASASSN-14li at the time of E1.

A double blackbody fitting to the UV-optical-IR SED at E1 
yields a dust temperature (\tdust) of $2091\pm172$~K,
contributing an integrated IR luminosity of $2.5\times10^{41}$~\lum.
The X-ray ($\sim7.7\times10^5$~K) and UV-optical temperatures ($\sim2.6\times10^4$~K) 
were nearly constant over a period of several months 
(Holoien et al. 2015; Cenko et al. 2016; van Velzen et al. 2016a) 
although their luminosities were declining.
The dust is quite hot, with a temperature even slightly higher than the typical
sublimation temperature (\tsub; e.g., 1400~K for silicate, 1800~K for graphite). 
The peak luminosity ($\sim10^{44}$~\lum) of ASASSN-14li determines a characteristic
dust sublimation radius (\rsub) of $\sim0.04$~pc\footnote{The formula we used to calculate
is that of footnote 1 in Namekata \& Umemura (2016).}, comparable with
the estimation dust location from the light-travel distance.
After heated by the initial continuum pulse, only grains outside the sublimation radius
at peak luminosity can survive. 
As a result, the high \tdust\ can be reasonably understood as
the IR emission we observed originates from the dust around the \rsub.
The same approach gives a \tdust of 588~K at E2, which is at $\sim210$ days 
after its discovery.

The dust mass associated with the IR emission can be also derived.
We assume that the dust grains follow a MRN size distribution
(Mathis, Rumpl, \& Nordsieck 1977; see also Draine \& Lee 1984) as 
$dn/da\propto a^{-3.5}$ with $a_{\rm min}=0.005\mu$m, $a_{\rm max}=0.3\mu$m
and an average density of $\rho=2.5\rm g~cm^{-3}$.
The total dust mass is estimated to be $\rm 1.2\times10^{27}g$, that is $\sim10^{-6}~\msun$.
Assuming a dust to gas ratio of 0.01, the amount of corresponding gas is $\sim10^{-4}~\msun$.

For a dust grain with a radius $a$ at a distance of $R$ from the heating source,
its equilibrium temperature $T$ is determined by the 
balance between the radiative heating by the absorption of UV-optical photons 
(proportional to the angular average of radiation intensity)
and the thermal re-emission in the IR.
Then the bolometric luminosity (\lbol) can be calculated by the equation below:

\begin{equation}
   \frac{L_{\rm bol}}{4\pi R^2} \pi a^2  = 4\pi a^2 \sigma T^4
\end{equation}
We adopt $R=c\Delta t$, in which $\Delta t=110$~days and $c$ is the light speed,
then $L_{\rm bol}=4.4\times10^{45}$~\lum, which is one order of magnitude higher 
than the observed peak luminosity. 
The blackbody assumption for the dust emission could be over-simplistic.
In a more practical calculation taking consideration of the absorption efficiency
($Q_{\rm abs}\propto \nu^{\beta}$, see formula 8.16 in Kruegel 2003),
the fitted \tdust=904~K (if we adopt $\beta=2$) and the required \lbol\ can 
be estimated as follows:
\begin{equation}
  \frac{L_{\rm bol}}{16\pi^2 R^2} \simeq 1.47\times10^{-6}aT^6
\end{equation}
The result is very sensitive to the grain size distribution:
$L_{\rm bol}=7.8\times10^{42}$~\lum\ assuming a MRN grain size distribution;
if we adopt a uniform grain size of $0.1\mu$m, $L_{\rm bol}$ turns out to be
$1.0\times10^{44}$~\lum\ (see relevant results in Table~\ref{wise}).
In this case, the estimated dust mass is $\sim10$~times higher ($\sim10^{-5}$~\msun)
than that assuming the simple blackbody radiation.

During our calculation above, we have assumed that the medium is 
optically thin, which is consistent with the fact that the infrared luminosity 
($\sim10^{41}$~\lum, see Table~1) is only a small fraction ($\sim1\%$) 
of the predicted luminosity if the dust intercept 
all of the available flux ($\sim10^{43}$~\lum, Lu et al. (2016)).
The derived \lbol\ above ($10^{43}-10^{45}$~\lum) is basically comparable 
to the observed peak luminosity of ASASSN-14li (Holoien et al. 2016), 
providing a substantial evidence of the IR echo associated with the TDE.
Moreover, the temperature at E2 (444 K) is roughly consistent
with the value predicted by $t^{0.5}$ decaying trend (494~K) under
the equilibrium assumption.

Previous works show that \oiii\ (\loiii$\sim10^{39}$~\lum) and radio emission 
do exist in PGC~043234 before the TDE, suggesting that the galaxy 
may host a low-luminosity AGN.
However, the lack of X-ray emission and the W1-W2 color of 
\wise\ imply that the nuclear activity is rather weak (Holoien et al. 2016).
The gas and dust we found from the dust echo of TDE
indicate that they are located quite close to the black hole prior 
to the disruption. It is interesting to investigate why the gas is there but is
not accreted efficiently by the black hole. 
Given the gas mass estimated above, a back of the envelope calculation suggests that 
Bondi accretion would yield an accretion rate of only $\sim10^{-7}$~\msun/year,
comparable to the amount required to feed the black hole as revealed 
by pre-disruption radio emission ($\sim10^{-4}$ Eddington ratio).
The other possible origin of the dust is from the unbound debris from the 
disrupted star itself either in the form of a wind, which is also thought 
to be responsible for the radio emission (Alexander et al. 2016),
or in the form of the unbound tail (Guillochon et al. 2016).
Assuming a velocity of $0.3c$ (Alexander et al. 2016), the distance that the 
unbound debris can travel is $\sim0.04~pc$, close to the \rsub.

This work demonstrates that the \wise\ and \neowise-R data afford us an excellent
opportunity to study the dust properties in the very vicinity of black holes residing 
in dormant (or weakly active) galaxies, by taking advantage of the dust 
echo light of TDEs (see also van Velzen et al. 2016b and Dou et al. 2016). 
Besides the studies of known TDEs, we can also attempt to discover new TDEs at the
infrared band by making full use of the released \wise\ and \neowise-R data, which have 
an all-sky covering, a wide time span ($\sim6$ years till now) and a unique survey cadence.
Our preliminary try shows that this method is quite efficient and 
some TDE candidates with similar IR echo signals can be chosen
in a blind search. In the future, we hope
to characterize the nuclear environments of inactive galaxies statistically 
by comparing the luminosity functions of TDEs found in the infrared and other bands 
(e.g., optical, X-ray).

\acknowledgements

We thank the anonymous referee for a thorough report and many constructive
comments that helped us improve this work. 
We thank Wenbin Lu for helpful discussion on the TDE models.
The work is supported by National Basic Research Program of China (grant No. 2015CB857005), 
the Strategic Priority Research Program "The Emergence of Cosmological Structures" 
of the Chinese Academy of Sciences (XDB09000000), NSFC (NSFC-11233002, 
NSFC-11421303, U1431229), the Fundamental Research Funds for the Central Universities
and CAS "Light of West China" Program (2015-XBQN-B-5).
This research makes use of data products from the
{\emph{Wide-field Infrared Survey Explorer}, which is a joint
project of the University of California, Los Angeles,
and the Jet Propulsion Laboratory/California Institute
of Technology, funded by the National Aeronautics and
Space Administration. This research also makes use of
data products from {\emph{NEOWISE-R}}, which is a project of the
Jet Propulsion Laboratory/California Institute of Technology,
funded by the Planetary Science Division of the
National Aeronautics and Space Administration.

\end{document}